\documentclass{amsart}
\makeatletter

\newtheorem{thm}{Theorem}[section]
\def\statetheorem{\@ifnextchar[{\@statetheorem}{\nr@statetheorem}}
\long\def\@statetheorem[#1]#2{\begin{thm}\label{#1}#2\end{thm}}
\long\def\nr@statetheorem#1{\begin{thm}#1\end{thm}}
\def\statetheorempf{\@ifnextchar[{\@statetheorempf}{\nr@statetheorempf}}
\long\def\@statetheorempf[#1]#2{\begin{thm}\label{#1}#2\end{thm}\proof}
\long\def\nr@statetheorempf#1{\begin{thm}#1\end{thm}\proof}

\newtheorem{lmma}{Lemma}[section]
\def\statelemma{\@ifnextchar[{\@statelemma}{\nr@statelemma}}
\long\def\@statelemma[#1]#2{\begin{lmma}\label{#1}#2\end{lmma}}
\long\def\nr@statelemma#1{\begin{lmma}#1\end{lmma}}
\def\statelemmapf{\@ifnextchar[{\@statelemmapf}{\nr@statelemmapf}}
\long\def\@statelemmapf[#1]#2{\begin{lmma}\label{#1}#2\end{lmma}\proof}
\long\def\nr@statelemmapf#1{\begin{lmma}#1\end{lmma}\proof}

\newtheorem{crlry}{Corollary}[section]
\def\statecorollary{\@ifnextchar[{\@statecorollary}{\nr@statecorollary}}
\long\def\@statecorollary[#1]#2{\begin{crlry}\label{#1}#2\end{crlry}}
\long\def\nr@statecorollary#1{\begin{crlry}#1\end{crlry}}
\def\statecorollarypf{\@ifnextchar[{\@statecorollarypf}{\nr@statecorollarypf}}
\long\def\@statecorollarypf[#1]#2{\begin{crlry}\label{#1}#2\end{crlry}\proof}
\long\def\nr@statecorollarypf#1{\begin{crlry}#1\end{crlry}\proof}

\def\note{\@ifnextchar[{\@note}{\@note[Note]}}
\def\@note[#1]{\par\medskip\noindent{\textbf{#1:} }}
\def\C{\mathbb{C}}

\def\bibref[#1]{\cite{#1}}

\let\real@bibitem\bibitem
\def\bibitem[#1]{\real@bibitem{#1}}

\makeatother

\def\M{\mathcal{M}}
\def\X{\mathcal{X}}
\def\ellv{\vec{\ell}}
\def\lamv{\vec{\lambda}}
\def\ev{\vec{e}}
\def\fv{\vec{f}}
\def\vv{\vec{v}}
\def\wv{\vec{w}}

\title{Nested Bethe Ansatz and Finite Dimensional Canonical Commutation Relationship}

\author{Alex Kasman}
\address{Department of Mathematics, College of Charleston, Charleston,
SC 29424, USA}

\begin{document}

\begin{abstract}
Recent interest in discrete, classical integrable systems has focused
on their connection to quantum integrable systems via the Bethe
equations.  In this note, solutions to the rational nested Bethe
ansatz (RNBA) equations are constructed using the 
``completed Calogero-Moser phase space'' of matrices which satisfy a
finite dimensional analogue of the canonical commutation relationship.
A key feature is the fact that the RNBA equations are derived only from this
commutation relationship and some elementary linear algebra.  The
solutions constructed in this way inherit continuous and discrete
symmetries from the CM phase space.
\end{abstract}

\maketitle

\section{Introduction}

Let $\M_n$ be the set of all pairs of $n\times n$ complex matrices
$(X,Z)$ which satisfy the
commutation relationship
\begin{equation}
\textup{rank}([X,Z]+I)=1\label{eqn:basic}
\end{equation}
where $I$ is the $n\times n$ identity matrix.  (In other words, this
says that $[Z,X]\approx I$, up to a rank one deformation.)
Then $\M_n$ forms the ``adelic Grassmannian'' or
``completed Calogero-Moser phase space'' that arises in
the study of integrable systems of mathematical physics
\bibref[Kbislin,KKS,R,W2] as well as in the study of properties of
rings of differential operators \bibref[BW, W1].  It is the purpose of
this note to establish a relationship between these matrices and the
integrable system 
\begin{equation}
\prod_{k=1}^n
\frac{(x_j^m-x_k^{m-1})(x_j^m-x_k^m+\eta)(x_j^m-x_k^{m+1}-\eta)}{(x_j^m-x_k^{m-1}+\eta)(x_j^m-x_k^m-\eta)(x_j^m-x_k^{m+1})}=-1\qquad
\forall 1\leq j\leq n
\label{eqn:nba}
\end{equation}
for functions $x_j^m$ ($1\leq j\leq n$) of the discrete ``time''
parameter $m$.
Note that equations
\eqref{eqn:nba} arise independently as the \textit{Bethe equations}
for the elementary energies of certain solvable quantum models and as
a discretization of known classical particle systems
\bibref[KLWZ,KWZ,NRK].

One \textit{philosophy} behind the recent interest in \textit{discrete}
integrable systems is that things seem to be simpler, and theorems
true for more elementary reasons, in these situations.  This can be
seen for example from the fact that the connection between the
commutation relationship \eqref{eqn:basic} and the system \eqref{eqn:nba}
is proved by only a few
elementary identities of linear algebra.

\section{Some basic identities in the case $\det X=0$}

Fix a choice of $(X,Z)\in\M_n$ such that $\det (X)=0$ and let $\ev$
and $\fv$ be vectors such that $\ev\cdot\fv^{\top}$ is the rank one
matrix $[X,Z]+I$.  We will use the notation $\widetilde{M}$ to denote the
matrix of cofactors of the matrix $M$.  (In particular, if $M$ is
invertible then $\widetilde{M}=\det(M)M^{-1}$.)  Any matrix with
determinant zero has a matrix of cofactors with rank at most one, so
once again we may specify that $\vv$ and $\wv$ are two vectors such
that $\widetilde{X}=\vv\cdot\wv^{\top}$.

It will be convenient for us to refer to the following functions
$$
p(\lambda):=\fv^{\top} \cdot \widetilde{(\lambda
I-Z)}\cdot \vv
$$
$$
q(\lambda):=\wv^{\top}\cdot \widetilde{(\lambda I-Z)}\cdot \ev$$
and the constants $\gamma:=(\wv^{\top}\cdot \ev)$ and 
$\mu:=(\fv^{\top}\cdot \vv)$.

\begin{lemma}{We have:
\begin{enumerate}
\item
$
\det [(\lambda I-Z)\cdot X+I]=\gamma p(\lambda),$
\item
$\det [X\cdot (\lambda I-Z)  - I] =-\mu q(\lambda),$
\item
$\displaystyle
\det [(\lambda_1 I-Z)\cdot X\cdot (\lambda_2
I-Z)+(\lambda_2-\lambda_1)I]
=(\lambda_2-\lambda_1)p(\lambda_1)q(\lambda_2).
$
\end{enumerate}}

By the commutation relationship \eqref{eqn:basic} one has that
$$
(\lambda I-Z)\cdot X+I= X\cdot (\lambda I-Z) +\ev\cdot \fv^{\top}.
$$
Then, it is a general fact that the determinant of a rank one
perturbation of a matrix is given by the formula
$$
\det(M+\ev\cdot \fv^{\top})=\det(M)+\fv^{\top}\cdot \widetilde{M}\cdot
\ev.
$$
Applying this fact gives the first identity above.  The others follow
from the same sort of argument.
\end{lemma}

\section{Nested Bethe Ansatz Equations}

For any
pair $(X,Z)\in\M_n$ define the function
$$
\tau_{X,Z}(\ellv,\lamv)=\tau(\ellv,\lamv):=\det(X+\sum_{i=1}^{\infty}
 \ell_i (\lambda_i I-Z)^{-1}) $$ where $\ellv=(\ell_1,\ell_2,\ldots)$
and $\lamv=(\lambda_1,\lambda_2,\ldots)$
($\ell_i,\lambda_i\in\C$, only a finite number of $\ell_i\not=0$). 
This is a polynomial $\tau$-function for the KP hierarchy  written in so-called \textit{Miwa
 variables}.  One could show that in any three of the variables
 $\ell_i$ this function satisfies the discrete Hirota equations
 \bibref[KLWZ].  However, here we will only observe that the linear
 algebra identities above show that the roots of $\tau$ satisfy the
 RNBA equations \eqref{eqn:nba}.

The RNBA equations 
involve only two variables: $x$ and $m$.  These can be taken to
correspond to any two pairs $(\ell_i,\lambda_i)$.  So, let us choose
any two values for $i$: $i_1$ and $i_2$.  We adopt the notation of
\bibref[KWZ], specifying a non-zero lattice spacing $\eta\in\C$, and
writing the function $\tau$ as $\tau^m(x)$ where $x=\eta \ell_{i_1}$
and $m=\ell_{i_2}\in\C$.

Note that $\tau^m(x)$ is a polynomial in $x$ (a special case of the
elliptic polynomials considered in \bibref[KWZ]).  Denote by $x_j^m$ the
roots of the polynomials $\tau^m(x)$ (listed in any order).  The
connection between these functions and equations \eqref{eqn:nba}
is an immediate consequence of the following factorization of
$\tau$ evaluated at specific values of its arguments:
\begin{lemma}{Choosing either the top or bottom sign at every choice,
we have the two factorization formulas:
$$\tau^{m\mp 1}(x_j^m\pm \eta)=\frac{1}{\gamma \mu}(\pm \lambda_{i_2}\mp \lambda_{i_1})
\tau^{m\mp 1}(x_j^m)\tau^m(x_j^m\pm \eta).$$}

Let us denote by $X'$ the matrix $$X'=X+\frac{x_j^m}{\eta}
(\lambda_{i_1} I -Z)^{-1} + m (\lambda_{i_2}-Z)^{-1}.$$
Then, since $\tau^m(x_j^m)=0$ we know that $\det X'=0$.  Also, it is
clear that $\textup{rank}([X',Z]+I)=1$.  Hence we may apply the results of the
lemma of the previous section to the function
\begin{eqnarray*}
\tau^{m-1}(x_j^m+ \eta) &=& \det(X'+(\lambda_{i_1}I-Z)^{-1}
-(\lambda_{i_2}I-Z)^{-1})\\
 &=& \frac{1}{\det(\lambda_{i_1}I-Z)(\lambda_{i_2}I-Z)}
\det((\lambda_{i_1}I-Z)\cdot X'\cdot
(\lambda_{i_2}-Z)-(\lambda_{i_2}-\lambda_{i_1})I)
\\
 &=& \frac{\lambda_{i_2}-\lambda_{i_1}}{\det(\lambda_{i_1}I-Z)(\lambda_{i_2}I-Z)}
p(\lambda_1)q(\lambda_2)
\\
 &=& \frac{\lambda_{i_1}-\lambda_{i_2}}{\mu\gamma}
\tau^m(x_j^m+\eta)\tau^{m-1}(x_j^m).
\end{eqnarray*} 
Note that we first used part 3 of the previous lemma to factor $\tau^{m-1}(x_j^m+\eta)$ into a
constant times the
polynomials $p(\lambda_1)$ and $q(\lambda_2)$ and then used parts 1 and 2 to view those as
$\tau$ evaluated at other points.  The other equation in which the lower sign is chosen at each place is
proved similarly.
\end{lemma}

\begin{corollary}{The functions $x_j^m$ satisfy the RNBA
\eqref{eqn:nba} equations.}
Using the last lemma, we can factor one term and recombine two other
terms in this three term product to deduce that
\begin{eqnarray*}
&&\tau^{m+1}(x_j^m)\tau^m(x_j^m-\eta)\tau^{m-1}(x_j^m+\eta) 
\\ 
&=& 
\tau^{m+1}(x_j^m)\tau^m(x_j^m-\eta)
\left(\frac{1}{\gamma\mu}(\lambda_{i_1}-\lambda_{i_2})\tau^{m-1}(x_j)\tau^m(x_j+\eta)\right)
\\
 &=&
-\left(\frac{\lambda_{i_2}-\lambda_{i_1}}{\gamma\mu}\tau^{m+1}(x_j^m)\tau^m(x_j^m-\eta)\right)\tau^{m-1}(x_j)\tau^m(x_j+\eta)\\
 &=& -\tau^{m+1}(x_j^m-\eta)\tau^{m-1}(x_j)\tau^m(x_j+\eta).
\end{eqnarray*}
So, in particular,
$$
\frac{\tau^{m+1}(x_j^m)\tau^m(x_j^m-\eta)\tau^{m-1}(x_j^m+\eta)}{\tau^{m+1}(x_j^m-\eta)\tau^{m-1}(x_j)\tau^m(x_j+\eta)}=-1.
$$
Expanding this equation by the formula $\tau^m(x)=c\prod(x-x_j^m)$
(for some constant $c$)
yields exactly the RNBA equations.
\end{corollary}

As a consequence, we get our main result:

\statetheorem{Let $(X,Z)\in\M_n$, then for any
$\lambda_1,\lambda_2\in\C$ ($\lambda_2$ not an eigenvalue of $Z$), the
$n$ eigenvalues of the matrix $$ \X(m):=-\eta X\cdot
(\lambda_1-Z)-m\eta (\lambda_2-Z)^{-1}\cdot (\lambda_1-Z) $$ satisfy
equations \eqref{eqn:nba}.  Note that $\X(m+1)-\X(m)$ is a constant
matrix and hence we have represented solutions of \eqref{eqn:nba} as a
free flow on matrices.}

\subsection{Remarks}

As a result of this construction, several symmetries of the solution
space manifest themselves.  In particular, in addition to the obvious
symmetries presented by translation in $\ellv$ (the ``commuting
flows'' which appear as translations $X\to X+f(Z)$ on $\M_n$), there
is also the ``dual'' flow (cf.\ \bibref[cmbis,Kbislin]) corresponding
to translations of the form $Z\to Z+f(X)$.  In addition, the
parameters $\lamv$ allow for continuous deformation of the solutions
and the involution $(X,Z)\to(X^{T},Z^{T})$ on $\M_n$ provides discrete
symmetry of the of solutions.  It would be of interest to observe
whether this symmetry has any special significance for this system.
Recall (cf. \bibref[cmbis,KR,Kbislin,R,W2]) that this same involution
is a linearizing map for Calogero-Moser type particle systems and a
``bispectral involution'' for the Lax operators of rational solutions
to the KP hierarchy \bibref[W1].

As we know a great deal about the structure of $\M_n$ \bibref[W2],
this may be useful in finding new features of the Bethe ansatz
equations.  For instance, note that the matrix $\X(m)$ corresponding
to the simplest $2\times2$ matrices satisfying \eqref{eqn:basic}:
$$
X=\left(\begin{matrix}0&1\\
0&0\end{matrix}\right) \qquad Z=\left(\begin{matrix}0&0\\ 1 & 0\end{matrix}\right)
$$
has eigenvalues
$$
\frac{- \eta
        \left( \lambda_1m + 
          \lambda_2
           \left( -1 + \lambda_1 + m \right)  \right) 
         \pm \sigma(\lambda_1,\lambda_2,m)}{2\lambda_2}
$$
with
$$
\sigma(\lambda_1,\lambda_2,m)= {\sqrt{{\eta}^2
         \left( -4\lambda_1\lambda_2m
            \left( -1 + \lambda_2 + m \right)  + 
           {\left( \lambda_1m + 
               \lambda_2
                \left( -1 + \lambda_1 + m \right) 
               \right) }^2 \right) }}
$$
So, in particular, in the case $\lambda_1=\lambda_2$, they become
simply
$$
\{-\eta (\lambda_1 -1 + m), -\eta m\},
$$
a sort of ``bound state'' in which the eigenvalues remain a distance
of $\eta(\lambda_1-1)$ units apart regardless of the value of the discrete
time parameter $m$.

That the manifold $\M_n$ should be related to solutions of the RNBA
equations is not a surprise.  In fact, one knows from previous results
that these matrices can be used to write $\tau$-functions, that when
written in Miwa form would satisfy the Hirota bilinear difference
equation closely related to the RNBA equations (cf.\
\bibref[KLWZ] and \bibref[W2]).  The point of the observations above
is merely that one is able to precisely determine the connection
between the commutation relationship \eqref{eqn:basic} and the
dynamical system \eqref{eqn:nba} without refering to any of these more
general results.  In particular, the results above provide a means for
demonstrating the role of the manifold $\M_n$ in integrable systems
using only elementary linear algebra, without any mention of
symplectic geometry or Hamiltonian dynamics (as was used in
\bibref[KKS]).  It might be of interest to determine to what extent
one may rederive the known results about Calogero-Moser systems in the
continuum limit using
only the linear algebra identities recalled here.

\note[Acknowledgements]  I am grateful to P. Wiegmann for introducing
me to these discrete equations and teaching me about them.
J. Harnad's contribution to this note, a key step in the proof of the
main result, is especially appreciated.   Thanks also to H.~Widom, M.~Gekhtman,
E.~Formanek and the referee for their assistance.

\end{document}